\documentclass[11pt,titlepage,twoside,leqno]{article}
\usepackage{latexsym}
\usepackage{amsfonts}
\usepackage{amsmath}
\usepackage{amssymb}

\makeatletter

\newcount\hour  \newcount\minutes  \hour=\time  \divide\hour by 60
\minutes=\hour  \multiply\minutes by -60  \advance\minutes by \time
\def\mmmddyyyy{\ifcase\month\or Jan\or Feb\or Mar\or Apr\or May\or Jun\or Jul\or
  Aug\or Sep\or Oct\or Nov\or Dec\fi \space\number\day, \number\year}
\def\hhmm{\ifnum\hour<10 0\fi\number\hour :%
  \ifnum\minutes<10 0\fi\number\minutes}
\def\Draft{{\it Draft of \mmmddyyyy}}

\def\ps@jtsheadings{%
\def\@oddhead{\it\rightmark\hfil\rm\thepage}%
\def\@oddfoot{\hfil\Draft}%
\if@twoside
\def\@evenhead{\rm\thepage\hfil\it\leftmark}%
\def\@evenfoot{\Draft\hfil}%
\else
\let\@evenhead\@oddhead%
\let\@evenfoot\@oddfoot%
\fi
}
\def\ps@jtsplain{%
\def\@oddhead{\hfil\Draft}%
\def\@oddfoot{\hfil\rm\thepage\hfil}%
\let\@evenfoot\@oddfoot%
\if@twoside \def\@evenhead{\Draft\hfil} \else \let\@evenhead\@oddhead \fi
}

\def\chaptermark#1{\markboth{\thechapter.\ #1}{\thechapter.\ #1}}%
\def\sectionmark#1{\markright{\thesection.\ #1}}

\def\section{\@startsection {section}{1}{\z@}
    {3.5ex plus1ex minus.2ex}{2.3ex plus.2ex}{\Large\bf}}
\def\subsection{\@startsection{subsection}{2}{\z@}
    {3.25ex plus1ex minus.2ex}{1.5ex plus.2ex}{\large\bf}}
\def\subsubsection{\@startsection{subsubsection}{3}{\z@}
    {3.25ex plus1ex minus.2ex}{1.5ex plus.2ex}{\normalsize\bf}}
\def\paragraph{\@startsection{paragraph}{4}{\z@}
    {3.25ex plus1ex minus.2ex}{1em}{\normalsize\bf}}
\def\subparagraph{\@startsection{subparagraph}{4}{\parindent}
    {3.25ex plus1ex minus.2ex}{1em}{\normalsize\bf}}

\makeatother

\makeatletter \@beginparpenalty=10000 \makeatother
\def\underl#1 {\leavevmode\let\first=\relax\underli #1 }
\def\underli#1 {\ifx&#1\let\next=\relax\unskip
                \else\let\next=\underli\first\ulinebox{#1}\fi\let\first=\undersp\next}
\def\undersp{\penalty50\ulinebox{\space}\penalty50}
\def\ulinebox#1{\vtop{\hbox{\strut#1}\hrule}}
\def\unice#1 {\underl #1 & }

\def\desclabel#1{\bf #1\hfil}
\def\desc{\list{}{%
\labelwidth=\leftmargin
\advance \labelwidth by -\labelsep
\let \makelabel=\desclabel}}

\makeatletter 

\newlength{\filength}
\newsavebox{\gcbox}
\sbox{\gcbox}{\framebox[\filength]{\rule{0ex}{2ex}}}

\newlength{\leftjustindent}
\newlength{\@leftjustindent}
\def\leftjust{\let\\\@leftjustcr\let\end\@endleftjust
  \addtolength{\@leftjustindent}{\leftjustindent}
  \vcenter\bgroup
  \halign\bgroup
    \hbox to\displaywidth{
      \rule{\@leftjustindent}{0ex}$\displaystyle##$\hfill
      }\crcr
}
\def\endleftjust{\crcr\egroup\egroup\endgroup}
\def\@endleftjust#1{\crcr\egroup\egroup\@checkend{#1}\endgroup}
\def\@leftjustcr{\crcr}

\newtheorem{theorem}{Theorem}[section]
\newtheorem{corollary}[theorem]{Corollary}
\newcommand{\qedblob}{\mbox{\rule[-1.5pt]{5pt}{10.5pt}}}
\def\literalqed{{\ \nolinebreak\hfill\mbox{\qedblob\quad}}}

\def\qed{\literalqed}

\newtheorem{lemma}[theorem]{Lemma}

\newtheorem{proposition}[theorem]{Proposition}
\newcommand{\singlespacing}{\let\CS=
\@currsize\renewcommand{\baselinestretch}{1}\tiny\CS}
\newcommand{\singlespacingplus}{\let\CS=
\@currsize\renewcommand{\baselinestretch}{1.25}\tiny\CS}
\newcommand{\doublespacing}{\let\CS=
\@currsize\renewcommand{\baselinestretch}{1.75}\tiny\CS}
\newcommand{\draftspacing}{\let\CS=
\@currsize\renewcommand{\baselinestretch}{2.0}\tiny\CS}
\newcommand{\foospacing}{\let\CS=
\@currsize\renewcommand{\baselinestretch}{1.05}\tiny\CS}

\makeatother

\mathcode`\0="0030      
\mathcode`\1="0031
\mathcode`\2="0032
\mathcode`\3="0033
\mathcode`\4="0034
\mathcode`\5="0035
\mathcode`\6="0036
\mathcode`\7="0037
\mathcode`\8="0038
\mathcode`\9="0039
\singlespacingplus

\newtheorem{definition}[theorem]{Definition}

\makeatletter
\clubpenalty=\@highpenalty
\widowpenalty=\@highpenalty
\makeatother

\makeatletter
\newcommand{\niceonespacing}{\let\CS=\@currsize\renewcommand{\baselinestretch}{1.1}\tiny\CS}\newcommand{\nicetwospacing}{\let\CS=\@currsize\renewcommand{\baselinestretch}{1.2}\tiny\CS}
\newcommand{\nicethreespacing}{\let\CS=\@currsize\renewcommand{\baselinestretch}{1.3}\tiny\CS}
\newcommand{\singlespacingplusplus}{\let\CS=\@currsize\renewcommand{\baselinestretch}{1.35}\tiny\CS}
\newcommand{\nicefourspacing}{\let\CS=\@currsize\renewcommand{\baselinestretch}{1.4}\tiny\CS}
\newcommand{\nicefivespacing}{\let\CS=\@currsize\renewcommand{\baselinestretch}{1.5}\tiny\CS}
\newcommand{\nicesixpacing}{\let\CS=\@currsize\renewcommand{\baselinestretch}{1.6}\tiny\CS}
\makeatother

\makeatletter
\def\@cite#1#2{[#1\if@tempswa , #2\fi]}
\makeatother

\makeatletter
\def\@citex[#1]#2{\if@filesw\immediate\write\@auxout{\string\citation{#2}}\fi
  \def\@citea{}\@cite{\@for\@citeb:=#2\do
    {\@citea\def\@citea{,\linebreak[0]}\@ifundefined
       {b@\@citeb}{{\bf ?}\@warning
       {Citation `\@citeb' on page \thepage \space undefined}}%
\hbox{\csname b@\@citeb\endcsname}}}{#1}}
\makeatother

\makeatletter
\def\ps@thesis{\def\@oddhead{\hfil\rm\thepage\hfil}\def\@oddfoot{}\def\@evenhead{\hfil\rm\thepage\hfil}\def\@evenfoot{}\def\chaptermark##1{}\def\sectionmark##1{}}
\makeatother

\makeatletter
\def\foobarpt{\textfont\z@\tenrm 
  \scriptfont\z@\ninrm \scriptscriptfont\z@\sevrm
\textfont\@ne\tenmi \scriptfont\@ne\ninmi \scriptscriptfont\@ne\sevmi
\textfont\tw@\tensy \scriptfont\tw@\ninsy \scriptscriptfont\tw@\sevsy
\textfont\thr@@\tenex \scriptfont\thr@@\tenex \scriptscriptfont\thr@@\tenex
\def\unboldmath{\everymath{}\everydisplay{}\@nomath\unboldmath
          \textfont\@ne\tenmi 
          \textfont\tw@\tensy \textfont\lyfam\tenly
          \@boldfalse}\@boldfalse
\def\boldmath{\@ifundefined{tenmib}{\global\font\tenmib\@mbi\@magscale1\global
        \font\tensyb\@mbsy \@magscale1\global\font
         \tenlyb\@lasyb\@magscale1\relax\@addfontinfo\@xiipt
              {\def\boldmath{\everymath
                {\mit}\everydisplay{\mit}\@prtct\@nomathbold
                \textfont\@ne\tenmib \textfont\tw@\tensyb 
                \textfont\lyfam\tenlyb\@prtct\@boldtrue}}}{}\@xiipt\boldmath}%
\def\prm{\fam\z@\tenrm}%
\def\pit{\fam\itfam\tenit}\textfont\itfam\tenit \scriptfont\itfam\ninit
   \scriptscriptfont\itfam\sevit
\def\psl{\fam\slfam\tensl}\textfont\slfam\tensl 
     \scriptfont\slfam\tensl \scriptscriptfont\slfam\tensl
\def\pbf{\fam\bffam\tenbf}\textfont\bffam\tenbf 
   \scriptfont\bffam\ninbf \scriptscriptfont\bffam\ninbf 
\def\ptt{\fam\ttfam\tentt}\textfont\ttfam\tentt
   \scriptfont\ttfam\nintt \scriptscriptfont\ttfam\nintt 
\def\psf{\fam\sffam\tensf}\textfont\sffam\tensf
    \scriptfont\sffam\tensf \scriptscriptfont\sffam\tensf
\def\psc{\@getfont\psc\scfam\@xiipt{\@mcsc\@magscale1}}%
\def\ly{\fam\lyfam\tenly}\textfont\lyfam\tenly 
   \scriptfont\lyfam\ninly \scriptscriptfont\lyfam\sevly
 \@setstrut \rm}

\makeatother

\newcommand{\parityp}{{\rm \oplus P}}

\newcommand{\p}{{\rm P}}

\newcommand{\np}{{\rm NP}}

\newcommand{\pp}{{\rm PP}}
\newcommand{\ep}{{\rm C\!\!\!\!=\!\!\!P}}
\newcommand{\smallep}{{\rm C\!\!=\!P}}
\newcommand{\bpp}{{\rm BPP}}

\newcommand{\conp}{{\rm coNP}}
\newcommand{\pspace}{{\rm PSPACE}}

\newcommand{\ph}{{\rm PH}}
\newcommand{\scriptph}{\mbox{\protect\scriptsize\rm PH}}

\singlespacingplus

\def\pair#1{{{\langle\!\!~#1~\!\!\rangle}}}

\newcommand{\sigmastar}{\mbox{$\Sigma^\ast$}}
\newcommand{\nats}{\mathbb{N}}

\newcommand\seq{\subseteq}

\newcommand\Lolra{\ \Longleftrightarrow \ }

\newcommand{\equalsdef}{\stackrel{\mbox{\protect\scriptsize df}}{=}}
\newcommand{\Oplus}{\mbox{\boldmath $\oplus$}}
\newcommand{\AND}{{\rm AND}}
\newcommand{\OR}{{\rm OR}}

\newcommand{\maj}{\mbox{\sc Maj}}
\newcommand{\parity}{\mbox{\sc Par}}
\newcommand{\equ}{\mbox{\sc Equ}}

\newenvironment{block}{\begin{list}{\hbox{}}{\leftmargin 1em
    \itemindent -1em \topsep 0pt \itemsep 0pt \partopsep 0pt}}{\end{list}}

\foospacing
\title{Immunity and Simplicity for Exact Counting and Other Counting
Classes }

\author{
J\"{o}rg Rothe\,\thanks{
\protect\singlespacing
Supported in part 
by grants
NSF-INT-9513368/\protect\linebreak[0]DAAD-315-PRO-fo-ab and
NSF-CCR-9322513 and
by a NATO Postdoctoral Science Fellowship
from the Deut\-scher Aka\-de\-mi\-scher Aus\-tausch\-dienst
(``Ge\-mein\-sames Hoch\-schul\-sonder\-pro\-gramm~III
von Bund und L\"andern'').
Work done in part while visiting the University of Rochester.
} 
\\ Institut f\"ur Informatik \\
Friedrich-Schiller-Universit\"at Jena \\
07740 Jena, Germany \\
{\tt rothe@informatik.uni-jena.de}
}

\date{Revision after Conference Final Version, May 12, 1998}

\makeatletter
\def\@listI{\leftmargin\leftmargini \parsep 4.5pt plus 1pt minus 1pt\topsep
6pt plus 2pt minus 2pt \itemsep  2pt plus 2pt minus 1pt}

\let\@listi\@listI
\@listi
\makeatother

\begin{document}

\typeout{WARNING:  BADNESS used to suppress reporting!  Beware!!}
\hbadness=3000
\vbadness=10000 

\bibliographystyle{alpha}

\pagestyle{empty}
\setcounter{page}{1}


\sloppy

\pagestyle{empty}
\setcounter{footnote}{0}

{\singlespacing

\maketitle

}


{\singlespacing
\begin{center}
{\large\bf Abstract}
\end{center}
\begin{quotation}
\vspace*{-1mm}
\noindent
Ko~\cite{ko:j:immune} and Bruschi~\cite{bru:j:immune} independently
showed that, in
some relativized world, PSPACE (in fact, $\parityp$) contains a set
that is immune to the polynomial hierarchy~(PH). In this paper, we
study and settle the question of (relativized) separations with
immunity for PH and the counting classes $\pp$, $\ep$, and $\parityp$
in all possible pairwise combinations. Our main result is that there
is an oracle $A$ relative to which $\ep$ contains a set that is immune
to~$\bpp^{\parityp}$.
In particular, this $\ep^A$ set is immune to $\ph^A$
and to~$\parityp^A$. Strengthening results of
Tor\'{a}n~\cite{tor:j:quantifiers} and Green~\cite{gre:j:threshold},
we also show that, in suitable relativizations, NP contains a
$\ep$-immune set, and $\parityp$ contains a $\pp^{\ph}$-immune set.
This implies the existence of a $\ep^B$-simple set for some
oracle~$B$, which extends results of Balc\'{a}zar et
al.~\cite{bal:j:simplicity,bal-rus:j:pro}, and provides the first
example of a simple set in a class not known to be contained in~PH.
Our proof technique requires a circuit lower bound for ``exact
counting'' that is derived from Razborov's~\cite{raz:j:majority} circuit
lower bound for majority.

\vspace*{2mm}
\noindent
{\bf Keywords:} Computational complexity; immunity;  counting classes;
relativized computation; circuit lower bounds.
\end{quotation}
}


%
%
\foospacing
\setcounter{page}{1}
\pagestyle{plain}
\sloppy

\clearpage

\section{Introduction}

A fundamental task in complexity theory is to prove separations or
collapses of complexity classes. Unfortunately, results of this kind
fall short for the most important classes between polynomial time and
polynomial space. In an attempt to find the reasons for this
frustrating failure over many years, and to gain more insight into why
these questions are beyond current techniques, researchers have
studied the problem of separating complexity classes in relativized
settings. Baker, Gill, and Solovay, in their seminal
paper~\cite{bak-gil-sol:j:rel}, gave for example relativizations $A$
and $B$ such that $\p^A \neq \np^A$ and $\p^B = \np^B$, setting the
stage for a host of subsequent relativization results.

Separations are also evaluated with regard to their quality. A {\em
simple separation\/} such as $\p^A \neq \np^A$ merely claims the
existence of a set $S$ in $\np^A$ that is not recognized by any $\p^A$
machine. This can be accomplished by a simple diagonalization ensuring
that every $\p^A$ machine fails to recognize $S$ by just one string,
which is put into the symmetric difference of $S$ and the machine's
language. It may well be the case, however, that some $\p^A$ machine
nonetheless accepts an infinite subset of~$S$, thus ``approximating
from the inside'' the set witnessing the separation. Thus, one might
argue that the difference between $\p^A$ and $\np^A$, as witnessed
by~$S$, is negligible. In contrast, a {\em strong separation\/} of
$\p^A$ and $\np^A$ is witnessed by a $\p^A$-immune set in~$\np^A$. For
any class ${\cal C}$ of sets, a set is {\em ${\cal C}$-immune\/} if it
is an infinite set having no infinite subset in~${\cal C}$. 

A relativization in which NP and P are strongly separated was first
given by Bennett and Gill~\cite{ben-gil:j:prob1}.
In fact, they prove a stronger result.
Technically speaking, they show that relative to a random oracle~$R$,
$\np^R$ contains a $\p^R$ bi-immune set with probability~1. This was
recently strengthened by Hemaspaandra and
Zimand~\cite{hem-zim:j:immunity} to the strongest result possible:
Relative to a random oracle~$R$, $\np^R$ contains a $\p^R$ {\em
balanced\/} immune set with probability~1. See these references for
the notions not defined here.  

Many more immunity results are known---see, e.g.,
\cite{hom-maa:j:oracle-lattice,boo-sch:j:imm,bal:j:simplicity,bal-rus:j:pro,tor-van:j:simplicity,bru-jos-you:j:strong,ko:j:immune,lis:j:immunity,bru:j:immune,epp-hem-tis-yen:j:mutual,bov-cre-sil:j:uniform,hem-rot-wec:j:easy}. Most
important for the present paper are the results and (circuit-based)
techniques of Ko~\cite{ko:j:immune} and
Bruschi~\cite{bru:j:immune}. In particular, both papers provide
relativizations in which the levels of the polynomial hierarchy (PH)
separate with immunity, Bruschi's results being somewhat stronger
and more refined, as they refer not only to the~$\Sigma$, but also to
the $\Delta$ levels of~$\ph$. Also, both authors independently obtain
the result that there exists a PH-immune set in PSPACE, relative to an
oracle. Since Ko's proof is only briefly sketched, Bruschi includes a
detailed proof of this result. This proof, however, is
flawed.\footnote{\protect\singlespacing In particular, looking into
the proof of~\cite[Thm.~8.3]{bru:j:immune}, the existence of the
desired oracle extension, $W$, in Case~(e) of the construction is not
guaranteed by the circuit lower bound used. In Case~(e) of Stage~$l$,
$W$ is required to have an odd number of length $h(l)$ strings such
that all circuits associated with a list of still unsatisfied
requirements reject their inputs {\em simultaneously\/}---an input
corresponds to the $W$ chosen; so once $W$ is fixed, every circuit has
the same input, $\chi_{W}(0^{h(l)}) \cdots \chi_{W}(1^{h(l)})$. The
used circuit lower bound for the parity function merely ensures that
for each circuit $C$ on that list, $C$ computes parity correctly for
at most 20\% of the ``odd'' inputs of length~$h(l)$. Thus, the
extension $W$ must be chosen according to the remaining 80\% of such
inputs to make that circuit reject. However, if there are sufficiently
many circuits on the list whose correct input regions happen to cover
{\em all\/} ``odd'' inputs of length~$h(l)$ (for instance, when there
are 5 circuits each being correct on a different 20\% of such inputs),
then there is no room left to choose a set $W \seq \{0,1\}^{h(l)}$ of
odd cardinality that makes all circuits reject simultaneously.}

Using Ko's approach, it is not difficult to give a valid and complete
proof of this result (and indeed the present paper provides such a
full proof---note Corollary~\ref{cor:imm}). However, the purpose of
this paper goes beyond that: We study separations with immunity for
counting classes inside PSPACE with respect to the polynomial
hierarchy and among each other.  Counting classes that have proven
particularly interesting and powerful with regard to the polynomial
hierarchy are $\pp$ (probabilistic polynomial time), the exact
counting class $\ep$, and $\parityp$ (parity polynomial time). Note
that the $\pspace^A$ set that is shown
by Ko~\cite{ko:j:immune} (cf.~\cite{bru:j:immune}) 
to be $\ph^A$-immune in fact is
contained in~$\parityp^A$. Ko's technique~\cite{ko:j:immune} is
central to all results of the present paper.

The relationship between these counting classes and PH still is a
major open problem in complexity theory, although surprising advances
have been made showing the hardness of counting. In particular,
Toda~\cite{tod:j:pp-ph} and Toda and
Ogihara~\cite{tod-ogi:j:counting-hard} have shown that each class
${\cal C}$ chosen among $\pp$, $\ep$, and $\parityp$ is hard for the
polynomial hierarchy (and, in fact, is hard for~${\cal C}^{\ph}$) with
respect to polynomial-time bounded-error random reductions.
Toda~\cite{tod:j:pp-ph} showed that $\pp$ is hard for $\ph$ even with
respect to deterministic polynomial-time Turing reductions. However,
it is widely suspected that PH is not contained in, and does not
contain, any of these counting classes. There are oracles known
relative to which each such containment fails, and similarly there are
oracles relative to which each possible containment for any pair of these
counting classes fails (except the known containment $\ep \seq \pp$
\cite{sim:thesis:complexity,wag:j:succinct}, which holds relative to
every oracle),
see~\cite{bak-gil-sol:j:rel,tor:thesis:count,tor:j:quantifiers,bei:j:mod,gre:j:threshold,bei:j:pp-oracle}.

Regarding relativized {\em strong\/} separations, however, the only
results known are the above-mentioned result that for some~$A$,
$\parityp^A$ contains a $\ph^A$-immune
set~\cite{ko:j:immune} (cf.~\cite{bru:j:immune}), 
and that for some~$B$, $\np^B$ (and
thus $\ph^B$ and~$\pp^B$) has a $\parityp^B$-immune
set~\cite{bov-cre-sil:j:uniform}. In this paper, we strengthen to
(relativized) strong separations all the other simple separations that
are possible among pairs of classes chosen from $\{\ph, \pp, \parityp,
\ep\}$.  Just as Balc\'{a}zar and
Russo~\cite{bal:j:simplicity,bal-rus:j:pro} exhaustively settled (in
suitable relativizations) all possible immunity and simplicity
questions among the probabilistic classes BPP, R, ZPP, and PP and
among these classes and P and~NP, we do so for the counting classes
$\ep$, $\pp$, and $\parityp$ among each other and with respect to the
polynomial hierarchy.

Ko's proof of the result that $\parityp^A$ contains a
$\ph^A$-immune set exploits the circuit lower bounds for the parity
function provided by Yao~\cite{yao:c:separating} and
H{\aa}stad~\cite{has:j:circuits}. Noticing that
H{\aa}stad~\cite{has:j:circuits} proved an equally strong lower bound
for the majority function, one could as well show that $\pp^A$
contains a $\ph^A$-immune set for some oracle~$A$. We prove a stronger
result: By deriving from Razborov's~\cite{raz:j:majority} circuit lower bound
for the majority function a sufficiently strong lower bound for the boolean
function that corresponds to ``exact counting,'' we construct an
oracle relative to which even in $\ep$ (which is contained in~PP)
there exists a set that is immune
even to the class $\bpp^{\parityp}$ (which contains PH by Toda's
result~\cite{tod:j:pp-ph}). This implies a number of new immunity
results, including (relativized) $\parityp$-immunity and
$\ph$-immunity of~$\ep$.

Conversely, we show that, in some relativized world, NP (and thus PH
and~PP) contains a $\ep$-immune set, which strengthens Tor\'{a}n's
simple separation of NP and
$\ep$~\cite{tor:thesis:count,tor:j:quantifiers}. As a corollary of
this result, we obtain that, in the same relativization, $\ep$ has a
{\em simple\/} set, i.e., a coinfinite $\ep$ set whose complement is
$\ep$-immune. Just like immunity, the notion of simplicity originates
from recursive function theory and has later proved useful also in
complexity theory. The existence of a simple set in a class ${\cal C}$
provides strong evidence that ${\cal C}$ separates from the
corresponding class~${\rm co}{\cal C}$.  Our result that, for some
oracle~$B$, $\ep^B$ has a
simple set extends Balc\'{a}zar's result that, for some~$A$,
$\np^A$ has a simple set~\cite{bal:j:simplicity}.  
We also strengthen to a strong
separation Green's simple separation that, relative to some oracle,
$\parityp \not\seq \pp^{\ph}$~\cite{gre:j:threshold}.  Similarly, the
(relativized) simple separation of the levels of the $\pp^{\ph}$
hierarchy~\cite{ber-ulf:jtoappear:perceptrons} also can be turned into
a strong separation. As a special case, this includes the existence of
a $\pp$-immune set in $\p^{\np}$ (and thus in~$\ph$) relative to some
oracle, which improves upon a simple separation
of Beigel~\cite{bei:j:pp-oracle}.

\section{Preliminaries}
\label{sec:prelims}

Fix the two-letter alphabet $\Sigma \equalsdef \{0,1\}$. The set of
all strings over $\Sigma$ is denoted~$\sigmastar$, and the set of
strings of length $n$ is denoted~$\Sigma^n$. For any string $x \in
\sigmastar$, let $|x|$ denote its length. For any set $L \seq
\sigmastar$, the complement of $L$ is $\overline{L} \equalsdef
\sigmastar \setminus L$, and the characteristic function of $L$ is
denoted by~$\chi_L$, i.e., $\chi_L(x) = 1$ if $x \in L$, and
$\chi_L(x) = 0$ if $x \not\in L$.  For the definition of relativized
complexity classes and of oracle Turing machines, we refer to any
standard text book on computational complexity (see, e.g.,
\cite{pap:b:complexity,bal-dia-gab:b:sctI,hop-ull:b:automata}).  For
any oracle Turing machine $M$ and any oracle~$A$, we denote the
language of $M^A$ by~$L(M^A)$, and we simply write $L(M)$ if $A =
\emptyset$. For classes ${\cal C}$ and ${\cal D}$ of sets, define
${\cal C}^{\mbox{\scriptsize ${\cal D}$}}$ to be $\bigcup_{D \in {\cal
D}} {\cal C}^D$, where ${\cal C}^D$ denotes the class of languages
accepted by ${\cal C}$ oracle machines with oracle~$D$.  For any
class~${\cal C}$, let ${\rm co}{\cal C}$ denote $\{ L \mid
\overline{L} \in {\cal C}\}$.
We use NPOTM as a shorthand for ``nondeterministic polynomial-time oracle
Turing machine.'' Let $\mbox{\rm acc}_{M^A}(x)$ 
(respectively, $\mbox{\rm rej}_{M^A}(x)$)
denote the number of accepting 
(respectively, rejecting) computation paths of NPOTM $M$ with
oracle $A$ on input~$x$, and let $\mbox{\rm tot}_{M^A}(x)$ be the total
number of computation paths of $M^A$ on input~$x$.
\begin{definition}
Let $A$ be any oracle set.
\begin{enumerate}
\item {\rm \cite{mey-sto:b:reg-exp-needs-exp-space,sto:j:poly}} \quad
The (relativized) polynomial hierarchy can be defined as follows, see
also~{\rm \cite{wra:j:complete}}:
\begin{itemize}
\item For each $k \geq 0$, a set $L$ is in $\Sigma_{k}^{p, A}$ if and
only if there exists a polynomial $p$ and a predicate $\sigma$
computable in $\p^A$ such that for all strings~$x$,
\begin{eqnarray*}
x \in L & \Longleftrightarrow & 
(\mbox{\rm Q}_1 w_1)\, (\mbox{\rm Q}_2 w_2)\, \cdots (\mbox{\rm Q}_k
w_k)\, [\sigma(x, w_1, w_2, \ldots , w_k)=1],
\end{eqnarray*}
where the $w_j$ range over the length $p(|x|)$ strings, and for
each~$i$, $1 \leq i \leq k$, $\mbox{\rm Q}_i = \exists$ if $i$ is odd, and
$\mbox{\rm Q}_i = \forall$ if $i$ is even. Let $\Pi_{k}^{p, A}$ denote
${\rm co} \Sigma_{k}^{p, A}$.  

\item Define $\ph^A \stackrel{\mbox{\protect\scriptsize\rm df}}{=}
\bigcup_{i \geq 0} \Sigma_{i}^{p, A}$.
\end{itemize} 

\item {\rm \cite{pap-zac:c:two-remarks,gol-par:j:ep}} \quad
$\parityp^A \stackrel{\mbox{\protect\scriptsize\rm df}}{=} \{ L \mid
(\exists\, \mbox{{\rm NPOTM} $M$})\, (\forall x \in \sigmastar)\, 
[x \in L \Lolra \mbox{\rm acc}_{M^A}(x) \mbox{\rm ~is odd}] \}$.

\item {\rm \cite{gil:j:probabilistic-tms}} \quad $\pp^A
\stackrel{\mbox{\protect\scriptsize\rm df}}{=} \{ L \mid (\exists\,
\mbox{{\rm NPOTM} $M$})\, (\forall x \in \sigmastar)\, 
[x \in L \Lolra \mbox{\rm acc}_{M^A}(x) \geq \mbox{\rm rej}_{M^A}(x)] \}$.

\item {\rm \cite{sim:thesis:complexity,wag:j:succinct}} \quad $\ep^A
\stackrel{\mbox{\protect\scriptsize\rm df}}{=} \{ L \mid (\exists\,
\mbox{{\rm NPOTM} $M$})\, (\forall x \in \sigmastar)\, 
[x \in L \Lolra \mbox{\rm acc}_{M^A}(x) = \mbox{\rm rej}_{M^A}(x)] \}$.

\item {\rm \cite{gil:j:probabilistic-tms}} \quad $\bpp^A$ is the class of
languages $L$ for which there exists an NPOTM $M$ such that
for each input~$x$, $x \in L$ implies that $\mbox{\rm rej}_{M^A}(x)
\leq \frac{1}{4}\mbox{\rm tot}_{M^A}(x)$, and $x \not\in L$ implies
that $\mbox{\rm acc}_{M^A}(x)\leq \frac{1}{4}\mbox{\rm tot}_{M^A}(x)$.

\item We write $\Sigma_{k}^{p}$ for $\Sigma_{k}^{p, \emptyset}$ and
$\ph$ for $\ph^{\emptyset}$, and similarly for the other classes.
\end{enumerate}
\end{definition}

Clearly, $\ph \cup \parityp \cup \pp \cup \ep \seq \pspace$ and $\bpp
\seq \pp$, and it is also known that $\bpp \seq \Sigma_{2}^{p} \cap
\Pi_{2}^{p}$~\cite{lau:j:bpp,sip:c:randomness} and $\conp \seq \ep
\seq \pp$~\cite{sim:thesis:complexity,wag:j:succinct}.

An $n$-ary boolean function is a mapping $f_n$ from $\{0,1\}^n$ to
$\{0,1\}$. Some of the most important boolean functions are the parity
function and the majority function. Let us define those functions that
will be considered in this paper:
\begin{itemize}
\item $\parity_n (x) = 1$ if and only if the number of bits of $x$ that
are 1 is odd.

\item $\maj_n (x) = 1$ if and only if at least $\lceil \frac{n}{2}
\rceil$ bits of $x$ are~1.

\item $\equ^{k}_{n} (x) = 1$ if and only if exactly $k$ bits of $x$
are~1, where $0 \leq k \leq n$.

\item $\equ^{\rm half}_{n} (x) = 1$ if and only if exactly $\lceil
\frac{n}{2} \rceil$ bits of $x$ are~1. 
\end{itemize}

Families of boolean functions are realized by circuit families. By
convention, when we speak of ``a'' circuit $C$ computing ``a''
function~$f$, we implicitly mean a family $C = (C_n)_{n \geq 0}$ of
circuits computing a family $f = (f_n)_{n \geq 0}$ of functions (i.e.,
for each~$n$, $C_n$ is a circuit with $n$ input gates and one output
gate that outputs the value $f_n (x)$ for each $x \in
\{0,1\}^n$). The {\em size\/} of a circuit is the number of its
gates. The {\em circuit complexity\/} (or {\em size\/}) of a boolean
function $f$ is the size of a smallest circuit computing~$f$. 
Unless stated otherwise, we will consider only constant depth,
unbounded fanin circuits with \AND, \OR, and $\Oplus$ (parity) gates. 
An \AND\ (respectively, \OR) gate outputs 1 (respectively, 0) if and only if
all its inputs are 1 (respectively, 0), and a $\Oplus$ gate outputs 1  
if and only if an odd number of its inputs are~1. 
Since $\{\AND, \OR, \Oplus\}$
(and indeed, $\{\AND, \Oplus\}$) forms a complete basis, we do not
need negation gates. Note that switching from one complete basis to
another increases the size of a circuit at most by a constant.  The
{\em depth\/} of a circuit is the length of a longest path from its
input gates to its output gate. Since adjacent levels of gates of the
same type can be collapsed to one level of gates of this type, we view
a circuit to consist of alternating levels of respectively \AND, \OR,
and $\Oplus$ gates, where the sequence of these operations is
arbitrary---the depth of the circuit thus also measures the number of
alternations.

\section{Immunity and Simplicity Results for Exact Counting}

In this section, we prove the main result of this paper: 

\begin{theorem}
\label{thm:imm}
There exists some oracle $A$ such that $\ep^{A}$ contains a
$\bpp^{\parityp^A}$-immune set.
\end{theorem}

Before turning to the actual proof, some technical details need be
discussed. First, we need a sufficiently strong lower bound on the
size of the ``exact counting'' function, $\equ^{\rm half}_{n}$, when
computed by circuits as described in the previous section. Razborov
proved the following exponential lower bound on the size of the
majority function when computed by such circuits
(see~\cite{smo:c:circuits} for a generalization of this result and a
simplification of its proof).

\begin{theorem}
\label{thm:majority}
{\rm \cite{raz:j:majority}} \quad For every~$k$, any depth $k$ circuit
with \AND, \OR, and $\Oplus$ gates that computes $\maj_n$ has size at
least $2^{\Omega(n^{1/(2k+2)})}$.
\end{theorem}

Using this lower bound for majority, we could (by essentially the same proof as
that of Theorem~\ref{thm:imm}) directly establish
$\bpp^{\parityp^A}$-immunity of~$\pp^A$. However, to obtain the
stronger result of Theorem~\ref{thm:imm}, we now derive 
from the above lower bound for majority a slightly
weaker lower bound for the $\equ^{\rm half}_{n}$ function, still being
sufficiently strong to establish Theorem~\ref{thm:imm}.

\begin{lemma}
\label{lem:equ}
For every~$k$, there exists a constant $\alpha_k > 0$ and an $n_k \in
\nats$ such that for all $n \geq n_k$, every depth $k$ circuit with
\AND, \OR, and $\Oplus$ gates that computes $\equ^{\rm half}_{n}$ has
size at least $n^{-1} \cdot 2^{\alpha_k n^{1/(2k+4)}}$.
\end{lemma}

\noindent
{\bf Proof.} Fix a sufficiently large~$n$. Clearly, the majority
function can be expressed as $\maj_n (x) = \bigvee_{i = \lceil
\frac{n}{2} \rceil}^{n} \equ_{n}^{i} (x)$. Each function
$\equ_{n}^{i}$, $0 \leq i \leq n$, is a subfunction of $\equ_{2n}^{\rm
half}$, since for each $x \in \{0,1\}^{n}$, $\equ_{n}^{i} (x) =
\equ_{2n}^{\rm half} (x0^i1^{n-i})$.  Thus, the circuit complexity of
$\equ_{n}^{i}$ is at most that of $\equ_{2n}^{\rm half}$ for
each~$i$. Now let $\mbox{size}_k (\equ_{n}^{\rm half})$ denote the
size of a smallest depth $k$ circuit with \AND, \OR, and $\Oplus$
gates that computes $\equ_{n}^{\rm half}$.  By the above observation,
we can realize $\maj_{\lceil \frac{n}{2} \rceil}$ with less than $n
\cdot \mbox{size}_k (\equ_{n}^{\rm half})$ gates in depth
$k+1$. Hence, by Theorem~\ref{thm:majority},
\[
\mbox{size}_k (\equ_{n}^{\rm half}) \geq n^{-1} \cdot  \mbox{size}_{k+1}
(\maj_{\lceil \frac{n}{2} \rceil}) = n^{-1} \cdot  2^{\alpha_k n^{1/(2k+4)}}
\]
for some suitable constant $\alpha_k > 0$ that depends on~$k$.~\qed

\medskip

For technical reasons, since we want to apply the above circuit lower
bound to obtain (relativized) $\bpp^{\parityp}$-immunity, we will now
give an equivalent definition of the class $\bpp^{\parityp}$ in terms
of a hierarchy denoted~$\ph^{\oplus}$.  As explained later,
$\ph^{\oplus}$ will only serve as a tool in the upcoming proof of
Theorem~\ref{thm:imm}. $\ph^{\oplus}$ generalizes the polynomial
hierarchy by allowing---in addition to existential and universal
quantifiers---the {\em parity\/} quantifier~{\boldmath $\Oplus$},
where $(\mbox{{\boldmath $\Oplus$}} w)$ means ``for an odd number of
strings~$w$.''

\begin{definition}
Let $A$ be any oracle set.
\begin{enumerate}
\item For each $k \geq 0$, a set $L$ is in $\ph_{k}^{\oplus, A}$ if
and only if there exists a polynomial $p$ and a predicate $\sigma$
computable in $\p^A$ such that for all strings~$x$, 
\begin{eqnarray*}
x \in L & \Longleftrightarrow &
(\mbox{\rm Q}_1 w_1)\, (\mbox{\rm Q}_2 w_2)\, \cdots (\mbox{\rm Q}_k w_k)\,
[\sigma(x, w_1, w_2, \ldots , w_k)=1],
\end{eqnarray*} 
where the $w_j$ range over the length $p(|x|)$ strings and the
quantifiers $\mbox{\rm Q}_j$ are chosen from $\{\exists, \forall,
\mbox{{\boldmath $\Oplus$}}\}$.

\item Define $\ph^{\oplus, A} \stackrel{\mbox{\protect\scriptsize\rm
df}}{=} \bigcup_{i \geq 0} \ph_{i}^{\oplus, A}$.

\item We write $\ph_{k}^{\oplus}$ for $\ph_{k}^{\oplus, \emptyset}$
and $\ph^{\oplus}$ for $\ph^{\oplus, \emptyset}$.
\end{enumerate}
\end{definition}

We stress that $\ph^{\oplus}$ is {\em not\/} a new complexity class or
hierarchy, since it is just another name for the
class~$\bpp^{\parityp}$, as can be proven by an easy induction from
the results of Toda~\cite{tod:j:pp-ph} and Regan and
Royer~\cite{reg-roy:j:error} that $\parityp^{\bpp^{\parityp}}$,
$\np^{\bpp^{\parityp}}$, and $\conp^{\bpp^{\parityp}}$ each are
contained in $\bpp^{\parityp}$.\footnote{\protect\singlespacing In
particular, due to these results, $\ph^{\oplus}$ in fact consists of
only four levels not known to be the same: $\ph^{\oplus}_{0} = \p$,
$\ph^{\oplus}_{1} = \np \cup \conp \cup \parityp, \ldots$,
and
$\ph^{\oplus}_{3} = \ph^{\oplus} = \bpp^{\parityp}$. Note also that
in~\cite{tod:j:pp-ph}, Toda preferred the operator-based notation,
which due to the closure of $\parityp$ under Turing reductions is
equivalent, i.e., ${\rm BP} \cdot \parityp = \bpp^{\parityp}$.
\label{foo:collapse} }
Rather, the purpose of $\ph^{\oplus}$ is merely to simplify the proof
of Theorem~\ref{thm:imm}. In particular, when using $\ph^{\oplus}$ in
place of~$\bpp^{\parityp}$, we do not have to deal with the promise
nature of $\bpp$ and, more importantly, we can straightforwardly
transform circuit lower bounds for constant depth circuits over the basis
$\{\AND, \OR, \Oplus\}$ into computations of $\ph^{\oplus}_{d}$ oracle
Turing machines.

Furst, Saxe, and Sipser~\cite{fur-sax-sip:j:parity} discovered the
connection between computations of oracle Turing machines and circuits
that allows one to transform lower bounds on the circuit complexity of
boolean functions such as parity into separations of relativized
PSPACE from the relativized polynomial hierarchy. (We adopt the
convention that for relativizing PSPACE, the space bound of the oracle
machine be also a bound on the length of queries it may ask, for
without that convention the problem of separating $\pspace^A$ from
$\ph^A$ becomes trivial,
see~\cite{fur-sax-sip:j:parity}.)~~Sufficiently strong (i.e.,
exponential) lower bounds for parity were then provided by
Yao~\cite{yao:c:separating} and H{\aa}stad~\cite{has:j:circuits}, and
were used to separate $\pspace^A$ from~$\ph^A$. They also proved lower
bounds for variations of the Sipser functions~\cite{sip:c:Borel} to
separate all levels of $\ph^A$ from each other (see
also~\cite{ko:j:exact}).

A technical prerequisite for this transformation to work is that the
computation of any $\Sigma_{i}^{p, A}$ machine can be simulated by a
$\Sigma_{i+1}^{p, A}$ machine that has the property that on all
computation paths at most one query is asked and this query is asked
at the end of the path (see~\cite[Cor.~2.2]{fur-sax-sip:j:parity}). An
oracle machine having this property is said to be {\em weak}. 
Similarly, the computation of any $\ph^{\oplus, A}_{i}$ machine can be 
simulated by a weak $\ph^{\oplus, A}_{i+1}$ machine.
The computation of a weak oracle machine $M^A$ on some input $x$ can then
be associated with a circuit whose gates correspond to the nodes of
the computation tree of $M^A(x)$, and whose inputs are the values
$\chi_{A}(z)$ for all strings $z \in \sigmastar$ that can be queried
by~$M^A(x)$. This correspondence can straightforwardly be extended to
the case of weak $\ph^{\oplus, A}$ oracle machines and is formally stated
in Proposition~\ref{prop:weak} below. The proof of
Proposition~\ref{prop:weak} is standard (see, e.g.,
\cite[Lemma~2.3]{fur-sax-sip:j:parity}
and~\cite[Lemma~2.1]{ko:j:exact} for analogous results) and thus
omitted. Let ${\cal CIR}(i,t)$ denote the collection of all depth
$i+1$ circuits with \AND, \OR, and $\Oplus$ gates, bottom fanin at
most~$t$, and fanin at most $2^t$ at all remaining levels.

\begin{proposition}
\label{prop:weak}
Let $A$ be any oracle and let $M$ be any weak $\ph^{\oplus , A}_{i}$
oracle machine running in time $p$ for some polynomial~$p$. Then, for
each $x \in \sigmastar$ of length~$n$, there exists a circuit
$C_{M,x}$ in ${\cal CIR}(i,p(n))$ whose inputs are the values of
$\chi_{A}(z)$ for all strings $z \in \sigmastar$ with $|z| \leq p(n)$
such that $C_{M,x}$ outputs 1 if and only if $M^{A}$ accepts~$x$. In
particular, it follows from the bounded depth and fanin of the
circuits in ${\cal CIR}(i,p(n))$ that the size of circuit $C_{M,x}$ is
bounded by $2^{s_{M}(n)}$ for some polynomial~$s_{M}$ depending
on~$M$.
\end{proposition}

Now we are ready to prove our main result.

\medskip

\noindent
{\bf Proof of Theorem~\ref{thm:imm}.} For any set~$S$, let
\[
L_{S} \equalsdef \{ 0^N \mid \mbox{$N \geq 1$ and the number of length
$N$ strings in $S$ equals $2^{N-1}$} \}.
\]
Clearly, for each~$S$, $L_{S}$ is in~$\ep^{S}$.

We will construct the set $A$ such that $L_{A} \in \ep^{A}$ is
$\ph^{\oplus , A}$-immune, i.e., $L_{A}$ is infinite and no infinite
subset of $L_{A}$ is contained in~$\ph^{\oplus , A}$. Since
$\bpp^{\parityp} = \ph^{\oplus}$ holds true in the presence of any
fixed oracle, this will prove the theorem. Also, since every
$\ph^{\oplus , A}_{d}$ machine can be transformed into a weak
$\ph^{\oplus , A}_{d+1}$ machine, it suffices to ensure in the
construction of $A$ that 
\begin{description}
\item[(a)] $L_{A}$ is infinite, and 
\item[(b)] for each
weak $\ph^{\oplus , A}$ oracle machine $M$ for which $L(M^{A})$ is an
infinite subset of~$L_{A}$, it holds that $M^{A}$ does not
recognize~$L_{A}$.
\end{description}

Fix an enumeration $M_{1}^{(\cdot)}, M_{2}^{(\cdot)}, \ldots$ of all
weak $\ph^{\oplus , (\cdot)}$ oracle machines; we assume the machines
to be clocked so that for each~$i$, the runtime of machine 
$M_{i}^{(\cdot)}$ is bounded by $p_{i}(n) = n^i + i$ for inputs
of length~$n$. 
In particular, if $i = \pair{d,j}$, the $i$th
machine $M_{i}^{(\cdot)}$ in this enumeration is the $j$th weak
$\ph^{\oplus , (\cdot)}_{d}$ oracle machine,
$M_{\pair{d,j}}^{(\cdot)}$, in the underlying enumeration
of weak $\ph^{\oplus, (\cdot)}_{d}$ oracle machines.  Satisfying
Property~(b) above then means to satisfy in the construction the
following requirement $R_i$ for each $i \geq 1$ for which $M_{i}^{A}$
accepts an infinite subset of~$L_{A}$:
\[
R_i :\quad L(M_{i}^{A}) \cap \overline{L_{A}} \neq \emptyset.
\]
We say that Requirement $R_i$ is {\em satisfied\/} if, at some point
in the construction of~$A$, $L(M_{i}^{A}) \cap \overline{L_{A}} \neq
\emptyset$ can be enforced.  

As a technical detail that is often used in immunity constructions, we
require our enumeration of machines to satisfy that for infinitely
many indices $i$ it holds that $M_{i}^{X}$ accepts the empty set for
every oracle~$X$, which can be assumed without loss of generality. We
will need this property in order to establish~(a).

Now we give the construction of~$A$, which proceeds in stages. In
Stage~$i$, the membership in $A$ of all strings up to length $t_i$
will be decided, and the previous initial segment of the oracle is
extended to~$A_i$.  Strings of length $\leq t_i$ that are not
explicitly added to $A_i$ are never added to the oracle. We define $A$
to be $\bigcup_{i \geq 0} A_i$.  Initially, $A_{0}$ is set to the
empty set and $t_{0} = 0$. Also, throughout the construction, we keep
a list ${\cal L}$ of unsatisfied requirements.  Stage $i > 0$ is as
follows.

\begin{description}
\item[Stage {\boldmath $i$}.] Add $i$ to~${\cal L}$. Consider all
machines $M_{\ell_1}^{(\cdot)}, \ldots , M_{\ell_m}^{(\cdot)}$
corresponding to indices $\ell_r$ that at this point are in~${\cal
L}$. Let $k = \max\{ d_r \mid \ell_r = \pair{d_r , j_r} \mbox{ and } 1
\leq r \leq m\}$ be the maximum level of the $\ph^{\oplus , (\cdot)}$
hierarchy to which these machines belong (not taking into account the
collapse of $\ph^{\oplus} = \bpp^{\parityp}$ mentioned in
Footnote~\ref{foo:collapse}).  Let $\alpha_{k+2} > 0$ be the constant
and $n_{k+2} \in \nats$ be the number that exist for depth $k+2$
circuits according to Lemma~\ref{lem:equ}. Choose $N = N_i >
\max\{t_{i-1}, \log n_{k+2}\}$ to be the smallest integer such that
\[
\alpha_{k+2} \cdot 2^{N/(2k + 8)} > N+i + \sum_{r = 1}^{m} s_{\ell_r}(N),
\]
where the polynomials $s_{\ell_r} = s_{M_{\ell_r}}$ correspond to the
machines with indices in ${\cal L}$ according to
Proposition~\ref{prop:weak}.

Distinguish two cases.
\begin{description}
\item[Case~1:] There exists an $r$, $1 \leq r \leq m$, and an
extension $E \seq \Sigma^{N}$ of $A_{i-1}$ such that $0^N \not\in L_E$
and yet $M_{\ell_r}^{A_{i-1} \cup E}$ accepts~$0^N$. Let $\tilde{r}$
be the smallest such~$r$. Cancel $\ell_{\tilde{r}}$ from ${\cal L}$,
set $A_i$ to $A_{i-1} \cup E$, and set $t_i$ to~$p_{i}(N)$.
Note that Requirement $R_{\ell_{\tilde{r}}}$ has been satisfied at
this stage.

\item[Case~2:] For all~$r$, $1 \leq r \leq m$, and for all extensions
$E \seq \Sigma^{N}$ of $A_{i-1}$, $0^N \not\in L_E$ implies that
$M_{\ell_r}^{A_{i-1} \cup E}$ rejects~$0^N$. In this case, no
requirement can be satisfied at this stage. However, to achieve
Property~(a), we will force $0^N$ into~$L_A$. Choose some extension
$\tilde{E} \seq \Sigma^{N}$ of $A_{i-1}$ such that (i)~the number of
length $N$ strings in $\tilde{E}$ equals $2^{N-1}$, and (ii)~for
each~$r$, $1 \leq r \leq m$, $M_{\ell_r}^{A_{i-1} \cup \tilde{E}}$
rejects~$0^N$. We will argue later (in Claim~1 below) that such an
extension $\tilde{E}$ exists. Set $A_i$ to $A_{i-1} \cup \tilde{E}$
and set $t_i$ to~$p_{i}(N)$.
\end{description}
\end{description}
{\bf End of Stage~{\boldmath $i$}.}

\medskip

Note that by the definition of $t_i$ and by our choice of~$N_i$, the
oracle extension in Stage~$i$ does not injure the computations
considered in earlier stages. Thus,  
\begin{eqnarray}
\label{equ:iff2}
(\forall i \geq 1) & & [0^{N_i} \in
L_{A_i}  \Lolra  0^{N_i} \in L_{A}] \mbox{, and} \\
\label{equ:iff3}
(\forall i, j \geq 1) & & [\mbox{$M_{j}^{A_{i}}$ accepts~$0^{N_i}$} 
 \Lolra  
\mbox{$M_{j}^{A}$ accepts~$0^{N_i}]$.}
\end{eqnarray}

The correctness of the construction will now follow from the following
claims.

\medskip

\noindent
{\bf Claim~1.} For each~$i \geq 1$, there exists an oracle extension
$\tilde{E}$ satisfying (i) and (ii) in Case~2 of Stage~$i$.

\medskip

\noindent
{\bf Proof of Claim~1.} Consider Stage~$i$. For each $r \in \{1,
\ldots , m\}$, let $C_{M_{\ell_r}, 0^N}$ be the circuit that, according
to Proposition~\ref{prop:weak}, corresponds to the computation of
$M_{\ell_r}$ running on input~$0^N$.  Fix all inputs to these circuits
except those of length $N$ consistently with~$A_{i-1}$. That is, for
each $r \in \{1, \ldots , m\}$, substitute in $C_{M_{\ell_r}, 0^N}$
the value $\chi_{A_{i-1}}(z)$ for all inputs corresponding to strings
$z$ with $|z| \leq t_{i-1}$, and substitute the value 0 for all inputs
corresponding to strings $z$ with $t_{i-1} < |z| \leq t_{i}$ and $|z|
\neq N$. Call the resulting circuits $\widehat{C}_{\ell_1, 0^N},
\ldots , \widehat{C}_{\ell_m, 0^N}$. By Proposition~\ref{prop:weak},
for each~$r$, $\widehat{C}_{\ell_r, 0^N}$ is in ${\cal
CIR}(k,p_{\ell_r}(N))$, its $2^N$ inputs correspond to the length $N$
strings, and for each $E \seq \Sigma^{N}$, it holds that
\begin{eqnarray}
\label{equ:iff1}
\mbox{$\widehat{C}_{\ell_r, 0^N}$ on input $\chi_{E}(0^N) \cdots
\chi_{E}(1^N)$ outputs 1} & \Longleftrightarrow &
\mbox{$M_{\ell_r}^{A_{i-1} \cup E}$ accepts~$0^N$.}
\end{eqnarray}

Create a new circuit $C_{2^N} = \OR_{r = 1}^{m} \widehat{C}_{M_{\ell_r},
0^N}$ whose $2^N$ inputs correspond to the length $N$ strings and
whose output gate is an \OR\ gate over the subcircuits
$\widehat{C}_{\ell_1, 0^N}, \ldots , \widehat{C}_{\ell_m, 0^N}$. Thus,
$C_{2^N}$ is a depth $k+2$ circuit with \AND, \OR, and $\oplus$ gates whose
size is bounded by 
\[
1 + \sum_{r = 1}^{m} 2^{s_{\ell_r}(N)} \leq 2^{i +
\sum_{r = 1}^{m} s_{\ell_r}(N)}
\] 
(note that $m \leq i$). 
By our choice of~$N$, we have $2^N
> n_{k+2}$ and
\[
2^{i + \sum_{r = 1}^{m} s_{\ell_r}(N)} < 2^{-N} \cdot 2^{\alpha_{k+2}
(2^N)^{1/(2k + 8)}}.
\]
Thus, by Lemma~\ref{lem:equ}, circuit $C_{2^N}$ cannot compute the
function $\equ^{\rm half}_{2^N}$ correctly for all inputs. Since by
the condition stated in Case~2 and by Equivalence~(\ref{equ:iff1})
above, $C_{2^N}$ behaves correctly for all inputs corresponding to any
set $E$ of length $N$ strings with $0^N \not\in L_E$, it follows that
$C_{2^N}$ must be incorrect on an input corresponding to some set
$\tilde{E}$ of length $N$ strings with $0^N \in L_{\tilde{E}}$, i.e.,
$C_{2^N}$ on input $\chi_{\tilde{E}}(0^N) \cdots \chi_{\tilde{E}}(1^N)$
outputs~0.  Since $C_{2^N}$ is the \OR\ of its subcircuits, each subcircuit
outputs 0 on this input. Thus, Equivalence~(\ref{equ:iff1}) implies
that for each~$r$, $1 \leq r \leq m$, $M_{\ell_r}^{A_{i-1} \cup
\tilde{E}}$ rejects~$0^N$.~\qed$_{\rm Claim~1}$

\medskip

\noindent
{\bf Claim~2.} $L_A$ is an infinite set.

\medskip

\noindent
{\bf Proof of Claim~2.} 
Recall our assumption that the index set of the empty set is infinite.
Since no requirement $R_i$ for which $i$ is an index of the empty set
can ever be satisfied and since, by construction, some
requirement is satisfied whenever Case~1 occurs, 
this assumption implies that Case~2 must happen
infinitely often. By construction, some string is forced into~$L_A$
whenever Case~2 occurs. Hence, $L_A$ is an infinite set. This proves
the claim and establishes Property~(a).~\qed$_{\rm Claim~2}$

\medskip

\noindent
{\bf Claim~3.} For every~$i \geq 1$, $M_{i}^{A}$ does not accept an
infinite subset of~$L_A$.

\medskip

\noindent
{\bf Proof of Claim~3.} For each~$i$, Requirement $R_i$ either is
satisfied at some stage of the construction, or is never satisfied.
If $R_i$ is satisfied at Stage~$j$, then Case~1 happens in Stage~$j$,
and so $0^{N_j} \in L(M_{i}^{A_{j}}) \cap \overline{L_{A_{j}}}$. By
Equivalences~(\ref{equ:iff2}) and~(\ref{equ:iff3}), $0^{N_j} \in
L(M_{i}^{A}) \cap \overline{L_{A}}$, so $L(M_{i}^{A}) \not\seq L_A$. 
Now suppose that Requirement
$R_i$ is never satisfied. We will argue that $L(M_{i}^{A}) \cap L_A$
then is a finite set. By construction, since we added to $A$ only
strings of lengths $N_j$, where $j \geq 1$ and $N_j$ is the integer
chosen in Stage~$j$, $L_A$ contains only strings of the form $0^{N_j}$
for some $j \geq 1$. Note that~$i$ is added to ${\cal L}$ in Stage~$i$
and will stay there forever. For each~$j \geq i$, if $0^{N_j} \in L_A$
(and thus $0^{N_j} \in L_{A_j}$ by~(\ref{equ:iff2})), then Case~2 must
have occurred in Stage~$j$. Consequently, $M_{i}^{A_{j}}$ (and thus
$M_{i}^{A}$ by~(\ref{equ:iff3})) rejects $0^{N_j}$ for every~$j \geq
i$. It follows that for each~$i$, $L(M_{i}^{A}) \cap L_A$ has at most 
$i-1$ elements, proving the claim.~\qed$_{\rm Claim~3}$

\medskip

Hence, $L_A$ is a $\bpp^{\parityp^A}$-immune set in~$\ep^{A}$.~\qed

\medskip

In particular, Theorem~\ref{thm:imm} immediately gives the following
corollary. All strong separations in Corollary~\ref{cor:imm} are new,
except the $\ph^A$-immunity of $\pspace^A$ (and of~$\p^{\pp^A}$, since
$(\forall B)\, [\parityp^B \seq \p^{\pp^B}]$), which is also stated
(or is implicit) in~\cite{ko:j:immune,bru:j:immune}, and except the
$\bpp^C$-immunity of $\pp^C$ (and its superclasses) proven
in~\cite{bal-rus:j:pro}. We also mention that 
Bovet et al.~\cite{bov-cre-sil:j:uniform} noted that $\pp^D$
strongly separates from $\Sigma_{2}^{p,D}$ for some oracle~$D$.

\begin{corollary}
\label{cor:imm}
Let ${\cal C}_1$ be any class chosen among $\ep$, $\pp$, 
$\p^{\smallep}$, $\p^{\pp}$, and $\pspace$, 
and let ${\cal C}_2$ be any class chosen among
$\bpp^{\parityp}$, $\bpp$, $\ph$, and $\parityp$.  There exists some oracle
$A$ such that ${\cal C}_{1}^{A}$ contains a ${\cal C}_{2}^{A}$-immune
set.
\end{corollary}

What about the converse direction? Does $\bpp^{\parityp}$, or even
some smaller class, contain a $\ep$-immune, or even a $\pp$-immune,
set relative to some oracle?  Note that
Tor\'{a}n~\cite{tor:thesis:count,tor:j:quantifiers} provided a simple
separation of this kind: There exists an oracle $A$ such that $\np^A
\not\seq \ep^A$ (see~\cite{bei:j:mod} for a simplification of the
proof of Tor\'{a}n's result). We strengthen this result by showing
that the separation is witnessed by a $\ep^B$-immune set in $\np^B$
for another oracle set~$B$. 
Indeed, the only property of $\ep$
needed to obtain a relativized separation from NP with immunity is that $\ep$
is closed under finite unions,\footnote{\protect\singlespacing It is
known that $\ep$ is closed even under polynomial-time ``positive''
Turing reductions, which is implicit in the methods
of~\cite{gun-nas-wec:c:counting-survey}, as has been noted
in~\cite{rot:t:some-closure} for the positive truth-table case; the
same result was noted independently
in~\cite{bei-cha-ogi:j:difference-hierarchies}. We refer to those
sources for a proof of Lemma~\ref{lem:ep}.  }
and this closure property relativizes.

\begin{lemma}
\label{lem:ep}
For every oracle~$A$, $\ep^A$ is closed under finite unions. That is,
given a finite collection $N_1, N_2, \ldots , N_k$ of NPOTMs, there
exists an NPOTM $N$ such that for each input~$x$, $N^A$ accepts $x$ (in
the sense of~$\ep$) if and only if for some~$j$, $N_{j}^{A}$
accepts~$x$ (in the sense of~$\ep$), i.e., for each $x \in \sigmastar$,
\[
\mbox{\rm acc}_{N^A}(x) = \mbox{\rm
rej}_{N^A}(x) \Lolra (\exists j : 1 \leq j \leq k)\, [\mbox{\rm
acc}_{N_{j}^{A}}(x) = \mbox{\rm rej}_{N_{j}^{A}}(x)] .
\]
\end{lemma}

\begin{theorem}
\label{thm:ep}
There exists some oracle $B$ such that $\np^B$ contains a
$\ep^B$-immune set.
\end{theorem}

\noindent
{\bf Proof.} 
The witness set here will be~$L_B$, where for any set~$S$, 
\[
L_{S} \equalsdef \{ 0^n \mid \mbox{$n \geq 1$ and there exists a
string of length $n$ in $S$} \}
\]
is a set in~$\np^{S}$. Fix an enumeration $N_{1}^{(\cdot)},
N_{2}^{(\cdot)}, \ldots$ of all NPOTMs, again
having the property that for infinitely many indices the machine with
that index accepts the empty set regardless of the oracle. (Throughout
this proof, ``acceptance'' means ``$\ep$ acceptance'' as in
Lemma~\ref{lem:ep}.) As in the proof of Theorem~\ref{thm:imm}, we try
to satisfy for each $i \geq 1$ for which $N_{i}^{B}$ accepts an infinite 
subset of~$L_{B}$, the requirement
\[
R_i :\quad L(N_{i}^{B}) \cap \overline{L_{B}} \neq \emptyset .
\]
Again, the stage-wise construction of $B = \bigcup_{i \geq
0} B_i$ is initialized by setting $B_{0}$ to the empty set and the
restraint function $t_{0}$ to~0, and we keep a 
list ${\cal L}$ of currently unsatisfied requirements.  
Stage $i > 0$ is as follows.

\begin{description}
\item[Stage {\boldmath $i$}.]  Add $i$ to~${\cal L}$. Consider all
machines $N_{\ell_1}^{(\cdot)}, \ldots , N_{\ell_m}^{(\cdot)}$
corresponding to indices $\ell_r$ that at this point are in~${\cal
L}$. Let $N^{(\cdot)}_{\mbox{\scriptsize $\cal L$}}$ be the machine that exists
for $N_{\ell_1}^{(\cdot)}, \ldots , N_{\ell_m}^{(\cdot)}$ by
Lemma~\ref{lem:ep}, i.e., for every oracle $Z$ and for each input~$x$,
\begin{eqnarray}
\label{equ:ep}
\mbox{$N^{Z}_{\mbox{\scriptsize $\cal L$}}$ accepts $x$} & 
\Longleftrightarrow & (\exists r : 1 \leq r \leq m)\, [\mbox{$N_{\ell_r}^{Z}$
accepts $x$}].
\end{eqnarray}
Let $p_{\mbox{\scriptsize $\cal L$}}$ be the polynomial bounding the
runtime of $N^{(\cdot)}_{\mbox{\scriptsize $\cal L$}}$. Choose $n = n_i
> t_{i-1}$ to be the smallest integer such that $2^n >
2p_{\mbox{\scriptsize $\cal L$}} (n)$.  Choose an oracle extension $E
\seq \Sigma^{n}$ of $B_{i-1}$ such that
\begin{eqnarray}
\label{equ:oracle}
E = \emptyset & \Longleftrightarrow & 
\mbox{$N^{B_{i-1} \cup E}_{\mbox{\scriptsize $\cal L$}}$ accepts~$0^n$.}
\end{eqnarray}
It has been shown in~\cite{bei:j:mod} that an oracle extension $E$
satisfying (\ref{equ:oracle}) exists if $n$ is chosen as above.
Set $B_i$ to $B_{i-1} \cup E$ and set $t_i$ 
to~$p_{\mbox{\scriptsize $\cal L$}}(n)$.
If the extension $E$ chosen is the empty set, then by (\ref{equ:oracle})
and (\ref{equ:ep}), there exists an~$r$, $1 \leq r \leq m$, 
such that $N_{\ell_r}^{B_{i-1}}$ accepts~$0^n$. Let $\tilde{r}$
be the smallest such~$r$, and cancel $\ell_{\tilde{r}}$ from~${\cal L}$.
\end{description}
{\bf End of Stage~{\boldmath $i$}.}

\medskip

Note that if we have chosen $E = \emptyset$ in Stage~$i$, then
$0^n \not\in L_E$ and Requirement $R_{\ell_{\tilde{r}}}$
has been satisfied. On the other hand, if $E \neq \emptyset$, 
then by (\ref{equ:oracle}) and (\ref{equ:ep}),  we have ensured that
(i)~$0^n \in L_E$, and (ii)~for each~$r$, $1 \leq r \leq m$, 
$N_{\ell_r}^{B_{i-1} \cup E}$ rejects~$0^n$. Now, an argument
analogous to Claims~2 and~3 in the proof of Theorem~\ref{thm:imm}
shows that $L_B$ is a $\ep^B$-immune set in~$\np^B$,
completing the proof.~\qed

\medskip

Similarly, there exists some oracle $C$ such that $\np^C$ (and
thus $\ph^C$ and~$\pp^C$) 
has a $\parityp^C$-immune set---this result was obtained
by Bovet et al.~\cite{bov-cre-sil:j:uniform}, based on their sufficient
condition for proving relativized strong separations and on Tor\'{a}n's
simple separation of $\np$ and
$\parityp$~\cite{tor:j:quantifiers}.  

Since the inclusions $\np \seq \pp$ and $\conp \seq \ep$ hold relative
to every fixed oracle, Theorem~\ref{thm:ep} immediately gives the
following corollaries.

\begin{corollary}
\label{cor:ep}
There exists some oracle $B$ such that $\pp^B$ contains a
$\ep^B$-immune set.
\end{corollary}

Recall from the introduction that for any complexity class~${\cal C}$,
a set is said to be {\em simple\/} for ${\cal C}$ 
(or {\em ${\cal C}$-simple\/}) if it
belongs to ${\cal C}$ and its complement is ${\cal C}$-immune. Homer
and Maass~\cite{hom-maa:j:oracle-lattice} proved the existence of a
recursively enumerable set $A$ such that $\np^A$ contains a simple
set, and Balc\'{a}zar~\cite{bal:j:simplicity} improved this result by
making $A$ recursive via a novel and very elegant trick: his
construction starts with a {\em full\/} oracle instead of an empty
oracle and then proceeds by {\em deleting\/} strings from it.
Balc\'{a}zar's result in turn was generalized by Torenvliet and van
Emde
Boas~\cite{tor:thesis:relativized-hierarchies,tor-van:j:simplicity} to
the second level and by Bruschi~\cite{bru:j:immune} to all levels of
the polynomial hierarchy. Balc\'{a}zar and Russo~\cite{bal-rus:j:pro}
also proved (relative to some oracle) the existence of a simple set 
in the one-sided error probabilistic class~R, which is contained in
$\np \cap \bpp$.  Our result below that $\ep$ has a simple set in some
relativization (all our oracles are recursive) extends those previous
simplicity results that each are restricted to classes contained in the
polynomial hierarchy.  Since of the classes we consider (PH, PP,
$\parityp$, and~$\ep$), all classes except $\ep$ are known to be
closed under complement, $\ep$ is the only class for which it makes
sense to ask about the existence of simple sets.

\begin{corollary}
There exists some oracle $B$ such that $\ep^B$ contains a simple set.
\end{corollary}

\noindent
{\bf Proof.} Let $B$ be the oracle constructed in the proof of
Theorem~\ref{thm:ep} and let $L_B$ be the witness set of this proof.
Consider the complement $\overline{L_B}$ of $L_B$ in~$\sigmastar$.
Since $L_B \in \np^B$, $\overline{L_B}$ is in $\conp^B$ and thus 
in~$\ep^B$. It has been shown in the proof of Theorem~\ref{thm:ep}
that~$L_B$, the complement of~$\overline{L_B}$, is an infinite set
having no infinite subset in~$\ep^B$. That is, $\overline{L_B}$  is 
$\ep^B$-simple.~\qed

\section{Immunity Results for {\boldmath $\parityp$} and 
the {\boldmath $\pp^{\ph}$} Hierarchy}

The last section in particular showed that, in suitable
relativizations, $\ep$ (and thus~$\pp$) is immune to both PH and
$\parityp$ (Corollary~\ref{cor:imm}), and NP (and thus PH and~PP) is
immune to $\ep$ (Theorem~\ref{thm:ep} and Corollary~\ref{cor:ep}) and
to $\parityp$~\cite{bov-cre-sil:j:uniform}.  In this section, we will
prove the existence of oracles
relative to which $\p^{\np}$ (and thus~PH) is
immune to~$\pp$, and relative to which $\parityp$ is immune
to~$\pp^{\ph}$.
The latter result strengthens the previously known (relativized)
strong separation of $\parityp$ from 
$\ph$~\cite{ko:j:immune} (cf.~\cite{bru:j:immune}),
and it also implies the new (relativized)
strong separation of $\parityp$ from~$\pp$.
Noticing that $\ep \seq \pp$ holds in all
relativizations, we thus have
settled all possible (relativized) strong separation questions
involving any pair of classes chosen among $\ph$, $\pp$, $\parityp$,
and~$\ep$, as claimed earlier.

We show these remaining results by
improving known (relativized) simple separations to strong ones. The
simple separation $(\exists A)\, [\parityp^A \not\seq
\pp^A]$~\cite{tor:thesis:count,tor:j:quantifiers} (see
also~\cite{bei:j:mod}) was strengthened by Green to $(\exists B)\,
[\parityp^B \not\seq \pp^{\ph^B}]$~\cite{gre:j:threshold}.

Since the analog of Lemma~\ref{lem:ep} as well holds for PP (in fact,
PP is closed under polynomial-time truth-table
reductions~\cite{for-rei:c:pp}, and this proof relativizes), 
the following theorem can be shown
by the technique used to prove Theorem~\ref{thm:ep}. 
First, we state the analog of
Lemma~\ref{lem:ep} in terms of weak $\pp^{\ph}$ oracle machines. The
proof of this lemma simply follows from the relativized version of the
proof that PP is closed under finite unions, which is a special case
of its closure under truth-table reductions~\cite{for-rei:c:pp}.

\begin{lemma}
\label{lem:pp}
Let $A$ be any oracle and $d \geq 0$ be any integer.  Given any finite
collection $N_1, N_2, \ldots , N_k$ of weak $\pp^{\ph}$ oracle
machines, there exists a weak $\pp^{\ph}$ oracle machine $N$ such that
for each input~$x$, $N^{A}$ accepts $x$ if and only if for some~$j$,
$1 \leq j \leq k$, $N_{j}^{A}$ accepts~$x$.
\end{lemma}

\begin{theorem}
\label{thm:pp}
There exists some oracle $D$ such that $\parityp^D$ (and thus
$\p^{\pp^D}$ and $\pspace^D$) contains a $\pp^{\ph^D}$-immune set.
\end{theorem}

\noindent
{\bf Proof.} Since the proof is very similar to that of Theorem~\ref{thm:ep},
we only mention the differences. 
The witness set here will be~$L_D$, where for any set~$S$, 
\[
L_{S} \equalsdef \{ 0^n \mid \mbox{$n \geq 1$ and there exists an odd
number of length $n$ strings in $S$} \}
\]
is a set in~$\parityp^{S}$. Now, $N_{1}^{(\cdot)}, N_{2}^{(\cdot)},
\ldots$ is an enumeration of all weak $\pp^{\ph^{(\cdot)}}$ oracle
machines, and ``acceptance'' refers to such machines. In Stage~$i$ of
the construction, we again consider all machines
$N_{\ell_1}^{(\cdot)}, \ldots , N_{\ell_m}^{(\cdot)}$ corresponding to
indices $\ell_r$ that at this point are in the list ${\cal L}$ of
currently unsatisfied requirements, and the machine
$N^{(\cdot)}_{\mbox{\scriptsize $\cal L$}}$ (with polynomial time
bound~$p_{\mbox{\scriptsize $\cal L$}}$) that exists for them by
Lemma~\ref{lem:pp}.  Assume $N^{(\cdot)}_{\mbox{\scriptsize $\cal L$}}$ is a
$\pp^{\Sigma_{d}^{p, (\cdot)}}$ machine, and let $c_d$ be the constant
that exists for such machines by~\cite[Thm.~5]{gre:j:threshold}.
Then, as shown in~\cite[Thm.~7]{gre:j:threshold}, choosing $n = n_i >
t_{i-1}$ to be the smallest integer such that
\[
2p_{\mbox{\scriptsize $\cal L$}}(n) \leq \min\{(2^n)^{1/d^2}, c_d 2^{n(d+1)/d^2}-1\}
\]         
implies that there exists an extension $E \seq \Sigma^{n}$ of the
oracle as constructed so far, $D_{i-1}$, such that $0^n \in L_E$ if
and only if $N^{D_{i-1} \cup E}_{\mbox{\scriptsize $\cal L$}}$
rejects~$0^n$.~\qed

\begin{corollary}
There exists some oracle $D$ such that $\parityp^D$ contains a
set immune to~$\pp^D$ and to~$\ph^D$.
\end{corollary}

By essentially the same arguments, also the very recent result
of Berg and Ulfberg~\cite{ber-ulf:jtoappear:perceptrons} that there is an 
oracle relative to which the levels
of the  $\pp^{\ph} = \bigcup_{d \geq 0} \pp^{\Sigma_{d}^{p}}$ hierarchy
separate (which generalizes Beigel's result that 
$(\exists A)\, [\p^{\np^A} \not\seq \pp^A]$~\cite{bei:j:pp-oracle}) 
can be strengthened to level-wise strong separations of this hierarchy. 
The proof of
Theorem~\ref{thm:perceptrons} is omitted, since it is very similar to
the previous proofs, the only difference being that it is based on the
construction given in~\cite{ber-ulf:jtoappear:perceptrons}. The
interested reader is referred to~\cite{rot:t:immunity} for a complete
proof of this result.

\begin{theorem}
\label{thm:perceptrons}
For any $d \geq 1$, there exists some oracle $F$ such that
$\p^{\Sigma_{d}^{p, F}}$ contains a $\pp^{\Sigma_{d-1}^{p, F}}$-immune
set. In particular, $\p^{\np^F}$ (and thus~$\ph^F$) has a
$\pp^F$-immune set.
\end{theorem}

\section{Conclusions and Open Problems}

In this paper, we have shown that all possible relativized separations
involving the polynomial hierarchy and the counting classes $\ep$,
$\pp$, and $\parityp$ can be made strong.  In particular, we have
extended to these counting classes previously known strong separations
of Ko~\cite{ko:j:immune} and Bruschi~\cite{bru:j:immune}, and we have
strengthened to strong separations previously known simple separations
of Tor\'{a}n~\cite{tor:thesis:count,tor:j:quantifiers},
Green~\cite{gre:j:threshold}, and Berg and
Ulfberg~\cite{ber-ulf:jtoappear:perceptrons}.  We have also shown
that $\ep$ contains a simple set relative to some oracle,
complementing the corresponding results of Balc\'{a}zar and
Russo~\cite{bal:j:simplicity,bal-rus:j:pro} for NP and~R, and of
Torenvliet and van Emde
Boas~\cite{tor:thesis:relativized-hierarchies,tor-van:j:simplicity}
and Bruschi~\cite{bru:j:immune} for $\Sigma_{k}^{p}$, $k > 1$.
However, many questions remain open. The most obvious question is
whether these immunity results can be strengthened to bi-immunity or
even to balanced immunity (see, e.g., \cite{hem-zim:j:immunity}).

Regarding the existence of simple sets in~$\ep^B$, note that our
construction of $B$ can easily be interleaved with other immunity
oracle constructions to show results such as: There exists an oracle
$A$ such that $\ep^A$ contains a simple set and another set that is
$\p^A$-immune (see~\cite{bal:j:simplicity} for the analogous result
for~NP). Torenvliet and van Emde
Boas~\cite{tor:thesis:relativized-hierarchies,tor-van:j:simplicity}
have even constructed an oracle relative to which NP contains a
language that {\em simultaneously\/} is simple and P-immune. Can this
also be shown to hold for~$\ep$?

Our main result that there exists some $A$ such that $\ep^A$ contains
a $\bpp^{\parityp^A}$-immune set is optimal in the sense that for all
oracles~$B$, $\ep^B$ clearly is contained in $\pp^B$ and thus
in~$\pp^{\parityp^B}$. However, it is also known that $\bpp^{\parityp}
\seq \mbox{\rm
Almost}[\parityp]$~\cite{tod-ogi:j:counting-hard,reg-roy:j:error},
where for any relativized class~${\cal C}$, $\mbox{\rm Almost}[{\cal
C}]$ denotes the class of languages~$L$ such that for almost all
oracle sets~$X$, $L$ is in ${\cal C}^X$~\cite{nis-wig:j:hard}. It is
an open problem (see~\cite{reg-roy:j:error}) whether $\bpp^{\parityp}
= \mbox{\rm Almost}[\parityp]$, so it is possible that $\mbox{\rm
Almost}[\parityp]$ is a strictly larger class than~$\bpp^{\parityp}$.
It is unlikely that $\ep$ is contained in $\mbox{\rm
Almost}[\parityp]$.  Is there an oracle relative to which $\ep$ is
even immune to $\mbox{\rm Almost}[\parityp]$? We conjecture that this
is the case.  Relatedly, can any of the immunity results of this paper
be shown to hold with probability 1 relative to a random oracle?

\bigskip

{\samepage
\noindent 
{\bf Acknowledgments.} \quad I am very grateful to Lane Hemaspaandra
for his constant and warm encouragement, for many incisive comments
and important suggestions that have much improved this paper, and for
careful proofreading. Interesting and helpful discussions with Gerd
Wechsung and Eric Allender are also acknowledged. I thank Christer
Berg and Staffan Ulfberg for providing me with an advance copy of
their paper~\cite{ber-ulf:jtoappear:perceptrons}.

}

{\singlespacing

}

\end{document}